\documentclass[prb,twocolumn,showpacs,preprintnumbers,amsmath,amssymb]{revtex4}
\usepackage{graphicx}
\usepackage{dcolumn}
\usepackage{bm}

\begin{document}


\title{Excitation mechanisms of individual CdTe/ZnTe quantum dots studied by photon correlation spectroscopy}

\author{J. Suffczy\'nski}
\email{Jan.Suffczynski@fuw.edu.pl}
\author{T. Kazimierczuk}
\author{M. Goryca}
\author{B. Piechal}
\author{A. Trajnerowicz}
\author{K.~Kowalik}
\author{P. Kossacki}
\author{A. Golnik}
\author{K. P. Korona}
\author{M. Nawrocki}
\author{J. A. Gaj}
\affiliation{Institute of Experimental Physics, Warsaw University, Ho\.za 69, 00-681 Warsaw, Poland}%

\author{G. Karczewski}
\affiliation{Institute of Physics, Polish Academy of Sciences, Al. Lotnik\'ow 32/64, 02-668 Warsaw, Poland}%

\date{\today}

\begin{abstract}
Systematic measurements of auto- and cross-correlations of photons
emitted from individual CdTe/ZnTe quantum dots under pulsed
excitation were used to elucidate non-resonant excitation mechanisms
in this self-assembled system. Memory effects extending over a few
excitation pulses have been detected in agreement with previous
reports and quantitatively described by a rate equation model,
fitting a complete set of correlation and PL intensity results. The
important role of single carrier trapping in the quantum dot was
established. An explanation was suggested for the unusually wide
antibunching dip observed previously in X-X autocorrelation
experiments on quantum dots under cw excitation.
\end{abstract}


\maketitle

\section{\label{sec:Intro}Introduction}

The emerging field of quantum information has given rise to an
interest for sources of photons-on-demand. Semiconductor light
sources offer important advantages, such as low power consumption
and possibilities of integration with existing electronics. Precise
knowledge of quantum dot (QD) excitation mechanisms is of primary
importance for the creation of semiconductor sources of
photons-on-demand. In CdTe/ZnTe QDs, some aspects of non-resonant
excitation are not yet fully understood, e.g., long components of
photoluminescence (PL) decay\cite{Piechal05} and the unusually wide
antibunching dip in autocorrelation of photons from exciton
recombination\cite{Couteau05} (excitonic photons). For the
prototypical InAs/GaAs system, it has been established by Santori et
al.\cite{SantoriSubm} that among the many memory processes in QD
optical excitation, medium time scale (sub$\mu$s) blinking leads to
the enhancement or suppression of excitonic photon autocorrelation
for resonant or non-resonant QD excitation respectively. The QD
charge state variation was suggested as the source of these effects.
We present here a systematic study of non-resonant excitation
mechanisms by photon correlation spectroscopy of CdTe/ZnTe single
quantum dots.

\section{\label{sec:Sample} Sample and characterization}

The sample was grown by molecular beam epitaxy on a (001) oriented
GaAs substrate, as described in Ref.~\onlinecite{Karcz99}. A
4~$\mu$m CdTe buffer layer was followed by a 100~nm ZnTe barrier.
Then two monolayers of CdTe were grown, forming fluctuation-type
quantum dots. The QD layer was overgrown by a ZnTe barrier of 50~nm
thickness. Transmission electron microscopy measurements performed
on such samples revealed quantum dots with a typical lateral size of
3~nm and a density of $10^{12}$~cm$^{-2}$
(Ref.~\onlinecite{Karcz99}).

\begin{figure*}
\includegraphics[height=55mm]{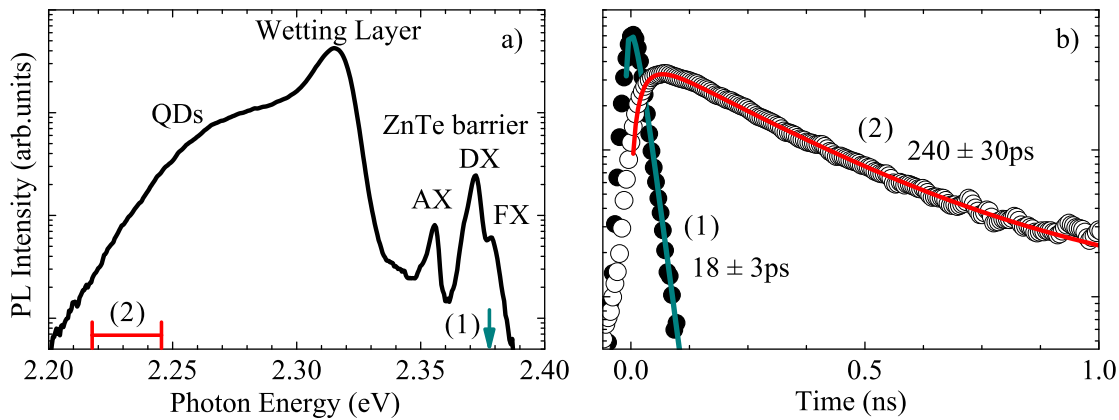}
\caption{\label{fig:ensamble}(Color online) (a) Time time-integrated
($0-1.6$~ns) PL spectrum of the CdTe/ZnTe sample at T = 7~K.
Structures originating from acceptor-bound (AX), donor-bound (DX),
and free (FX) excitons from ZnTe barrier as well as from excitons in
the wetting layer are indicated. Emission from QD layer forms a wide
band centered at 2.25~eV. Markers (1) and (2) indicate energies
corresponding to temporal profiles presented in
Fig.~\ref{fig:ensamble}(b). (b) PL transients of free excitons in
ZnTe barriers (1) and quantum dots (2) at T = 7~K. Points -
experimental data, solid lines - fitted curves. Respective decay
times indicated.}
\end{figure*}

The basic characterization was performed by standard CW
photoluminescence (PL) excited with an Ar-ion laser and by
time-resolved photoluminescence excited at $3.5$~eV by
frequency-doubled 150~fs pulses from a Ti:sapphire laser. In the
time resolved characterization a streak camera (resolution of about
10~ps) was used to record spectral and temporal distribution of PL.
A typical PL spectrum is presented in Fig.~\ref{fig:ensamble}(a).
Three major components can be distinguished. The set of three lines
at the highest energy ($2.345-2.385$~eV) is related to the exciton
emission from the barrier material (ZnTe). The middle energy
component ($2.30-2.34$~eV) is due to the wetting layer, and the one
at the lowest energy ($2.20-2.30$~eV) is related to the emission
from quantum dots. This natural assignment is consistent with the
results of further experiments.

The typical temporal profiles of the barrier and QD lines are
presented in Fig.~\ref{fig:ensamble}(b). The ZnTe emission consists
of a fast-decaying ($\tau_{D1}= 18 \pm 3$~ps) free exciton line at
$2.375 - 2.385$~eV and long-lived lines of acceptor-bound and
donor-bound excitons. The fast decay provides information on the
decay of free excitons and was used as an approximate measure of the
characteristic time of quantum dot excitation.
The decays of the QD PL lasted much longer. The dominant decay time
was about  $\tau_{D2}=240\pm30$~ps, but longer components were also
observed. The decay is not straightforward in interpretation. The
macro-PL signal is composed of many different lines related to
different quantum dots. In particular, lines due to neutral excitons
(X), biexcitons (XX) and trions (CX) may decay with different
characteristic times. Thus, the measurements on single quantum dots
were used to determine decay times of particular lines (see
Sec.~\ref{sqds.Decays}). However, the rise times of the QD PL
($\tau_R = 25 - 30$~ps) were significantly longer than the
resolution-limited rise times of the ZnTe PL ($\tau_R$ about 10 ps)
and close to the decay time of the FX line
(Fig.~\ref{fig:ensamble}(b)). This shows that the QDs are not
excited directly by light pulses and points to the transfer of
carriers (excitons) from barriers as a source of QD excitation. The
characteristic time of this transfer will be used in the rate
equation model presented in Sec.~\ref{modeldes.1}. The same values
have been determined by Viale et al.\cite{Viale03} in a PL dynamics
study of a similar CdTe/ZnTe QD system, and attributed to relaxation
processes from excited to the ground QD exciton state. As we will
see in Sec.~\ref{modeldes.2}, such an interpretation cannot be
maintained in our case.
\section{Single quantum dot spectroscopy}
In order to achieve a better insight in QD excitation and
recombination processes, several different experiments on single QDs
were performed such as precise determination of excitonic line
energy positions, measurement of QD in-plane anisotropy (not shown),
and dependence of QD PL intensity on excitation power
(Sec.~\ref{sqds.Ident}). In order to obtain excitonic radiative
lifetimes we measured the decays of individual QD emission lines
(Sec.~\ref{sqds.Decays}). Preliminary Excitation Correlation
Spectroscopy (CES) experiments on individual quantum dots were
performed (Sec.~\ref{sqds.ECS}) to estimate the temporal scale of
the QD excitation processes. Correlated photon counting with
femtosecond pulsed excitation was also performed
(Sec.~\ref{sqds.Correlations}). Both autocorrelation and
cross-correlation were measured, providing information on QD
excitation mechanisms.

\subsection{\label{sqds.Exp}Experiment}
For studies of individual quantum dots, a micro-photoluminescence
($\mu$-PL) setup was used, assuring a spatial resolution better than
$1\mu$m. The sample was fixed directly on the front surface of a
mirror type microscope objective\cite{Jasny} (N.A.$= 0.7$) inside a
pumped helium cryostat and cooled down to $T=1.7$~K. An Argon ion
laser (488~nm or multi-line UV) was used for cw excitation. For
time-resolved measurements, frequency-doubled pulses of a
mode-locked Ti:Sapphire laser were used (repetition frequency of
76~MHz, spectral and temporal width 2.6~nm and less than 1~ps
respectively). Averaged excitation power was varied in a range from
10~nW to 4~$\mu$W by use of neutral density filters.

In Excitation Correlation Spectroscopy the PL was excited by pairs
of pulses obtained by splitting of the pulsed laser beam. Temporal
separation between the two pulses of each pair, controlled by an
optical delay line, ranged from 0 to 1.5~ns. Time-integrated spectra
were recorded with a CCD camera. Single photon correlation
measurements were performed in a Hanbury-Brown and Twiss setup with
spectral filtering by grating monochromators (spectral resolution
200~$\mu$eV). Each monochromator was equipped with a CCD camera and
an avalanche photodiode (APD) serving as a single photon detector
(temporal resolution 750~ps, quantum efficiency 55\% at 560~nm, dark
counts $<$~200/s). For PL decay measurements one of the APDs was
replaced by a microchannel plate (MCP) photomultiplier tube
(temporal resolution of 40~ps). A correlation card with a
multichannel analyzer (4096 time bins, 146~ps width each) was used
to generate histograms of correlated photon detection events versus
delay, equivalent to an unnormalized second order correlation
function.

The same card was used to record PL decay curves by measuring time
distance between photon detection and laser pulse reference signal.
The channel width was set to 37~ps in that case.

Certain experiments were performed at doubled pulse repetition
frequency. This was achieved by dividing the laser beam in a $50/50$
nonpolarizing beamsplitter and delaying one of the components by
half (6.6~ns) of the initial repetition period, before combining
both components in a single beam.

\subsection{\label{sqds.Ident}Identification of single QD PL lines}
\begin{figure}
\includegraphics[height=65mm]{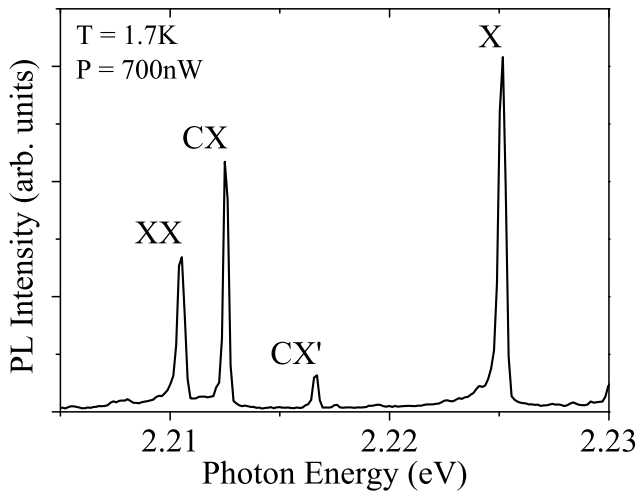}
\caption{\label{fig:PLspectrum} Photoluminescence spectrum of the
selected quantum dot, excited with frequency-doubled pulsed beam of
Ti:Sapphire laser at 402~nm (3~eV). The averaged excitation power
was 700~nW.}
\end{figure}

Because of the high dot density, individual dot lines could be well
resolved only in the long-wavelength tail of the spectrum. In a
previous study on the same sample\cite{Kudelski05} we found that a
typical spectrum of an individual QD contains a neutral exciton
line, accompanied by biexciton and charged exciton lines, at about
13~meV and 11~meV below respectively. The identification was
evidenced by a series of experiments, including synchronized line
energy jumps, in-plane optical anisotropy, and Zeeman effect
measurements. We selected for further measurements a quantum dot
emitting a group of PL lines presented in Fig.~\ref{fig:PLspectrum}.
The excellent mechanical stability of our experimental setup allowed
us to follow the microphotoluminescence of this quantum dot during
weeks of measurements. The main results of the paper were confirmed
for several other quantum dots. Besides the lines corresponding to
those identified in Ref.~\onlinecite{Kudelski05}, a weak line CX'
appears in the spectrum. Its relative energy position corresponds to
a negatively charged exciton line, identified by Besombes et
al.\cite{Besombes02} in photoluminescence of a similar system of
self assembled CdTe/ZnTe QDs. We assume therefore a tentative
identification of CX and CX' lines as due to positively and
negatively charged excitons respectively, recombining in the same
quantum dot.

\begin{figure}
\includegraphics[height=65mm]{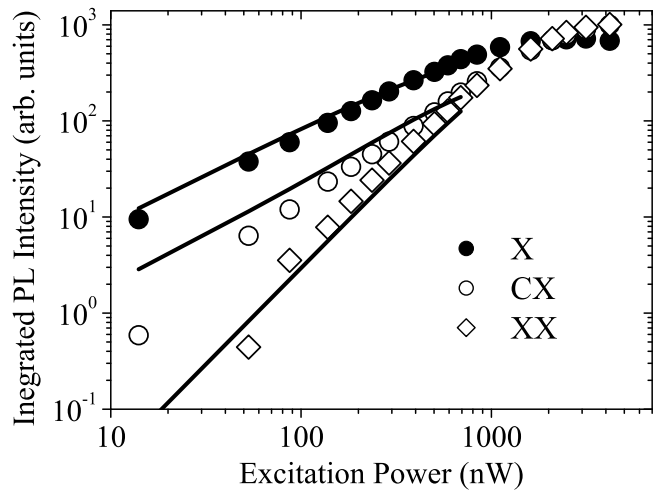}
\caption{\label{fig:PowerDep} Integrated photoluminescence intensity
at $T=1.7$~K as a function of excitation power for X, CX and XX.
Solid lines show model calculation (see Sec.~\ref{modeldes.2}).}
\end{figure}

By measuring in-plane optical anisotropy, we confirmed equal
absolute values and opposite signs of the anisotropic exchange
splittings for X and XX lines and no measurable anisotropy for CX
and CX' lines, as expected for the assumed identification
\cite{Kudelski05}. In further experiments we focused our attention
on the three strongest lines of the spectrum: X, CX, and XX. We
measured the PL spectra at different excitation powers.
Figure~\ref{fig:PowerDep} shows the power dependence of the
integrated intensity of each of the three lines. The observed
superlinear behavior of XX line and approximately linear one of X
and CX lines support the assumed identification.

Final proof of this identification is provided by single photon
correlation measurements presented below.

\subsection{\label{sqds.Decays}Measurements of PL decay on individual QD}
\begin{figure}
\includegraphics[height=100mm]{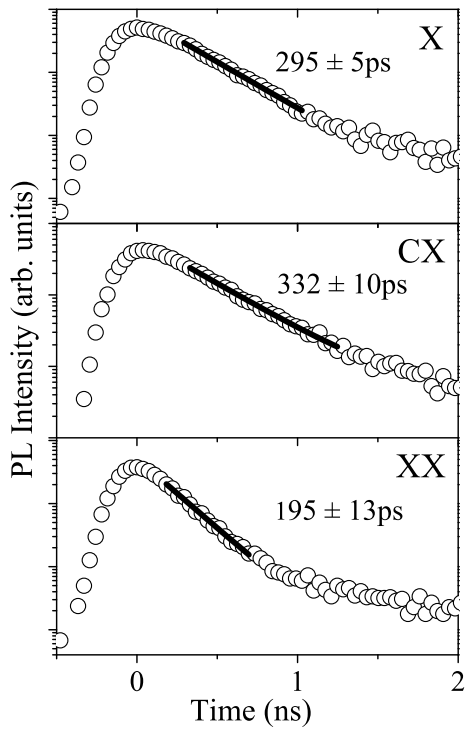}
\caption{\label{fig:Decays} Intensity of X, CX and XX
photoluminescence as a function of time following excitation pulse.
Solid lines represent exponential decays with indicated lifetime
values. Excitation power 700~nW, temperature 1.7~K.}
\end{figure}

The time-dependent intensities of PL emission from X, CX, and XX
states under pulsed excitation are presented in
Fig.~\ref{fig:Decays}.
Two components, a fast and a slow one, are present in all the
recorded decays, in agreement with previous observations on
CdTe/ZnTe QDs.\cite{Couteau05,Piechal05} The fast component
represents excitonic radiative lifetime, while the (much weaker)
slow component, usually in II-VI systems attributed to dark exciton
contribution,\cite{Labeau03,Patton03} is probably due to excitation
delayed by some trapping processes.\cite{Piechal05} The data was
fitted with monoexponential decay applied to the 'fast' part of the
decay curve (as indicated in Fig.~\ref{fig:Decays}), producing
radiative lifetime values $\tau_X = 295\pm5$~ps, $\tau_{CX} =
332\pm10$~ps, $\tau_{XX} = 195\pm13$~ps, for exciton, charged
exciton and biexciton respectively. The ratio of the exciton decay
time to the biexciton decay time is equal to 1.5, which is
consistent with previous results obtained on
CdTe/ZnTe\cite{Couteau05} and InAs/GaAs\cite{Santori02} QDs. The
decay time values obtained were used in the analysis of subsequent
experiments.

\subsection{\label{sqds.ECS}Excitation Correlation Spectroscopy}
\begin{figure}
\includegraphics*[width=85mm]{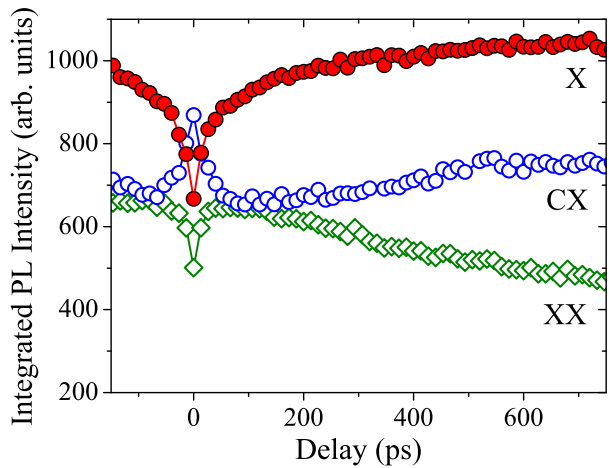}
\caption{\label{fig:2pulses} (Color online) Integrated intensity of
X, CX, and XX lines versus delay between two excitation pulses in
ECS experiment. Excitation power of a single beam 230~nW,
temperature 1.7~K.}
\end{figure}

The ECS technique, used mainly for studies of quantum wells,
\cite{Yamada95} has previously been applied to QDs to study coherent
processes.\cite{Besombes03,Young02} It provides an excellent
temporal resolution, limited principally by the width of the
exciting laser pulse. We performed preliminary ECS experiments to
check the time scale of QD excitation processes, inaccessible in our
single QD PL decay measurements. Figure~\ref{fig:2pulses} shows
integrated intensity of the three PL lines from a selected QD as a
function of the delay between the two excitation pulses. Besides the
effects on the scale of the radiative recombination times (hundreds
of ps), pronounced sharp structures are visible on a scale of order
of tens of ps. We attribute them to trapping of carriers and
excitons by the QD, in agreement with the discussion of time
resolved measurements on the QD ensemble in Sec.~\ref{sec:Sample}. A
detailed discussion of ECS measurements will be presented elsewhere.

\subsection{\label{sqds.Correlations}Single photon correlation measurements}
\begin{figure*}
\includegraphics{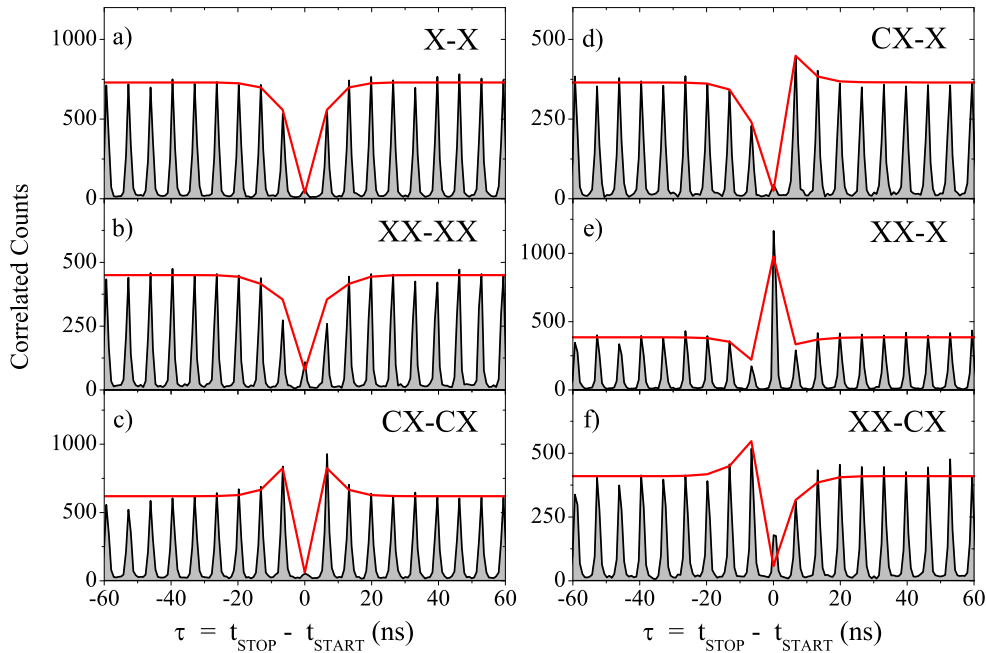}
\caption{\label{fig:histograms} (Color online) Histograms of
correlated counts as a function of time interval
$\tau=\tau_{STOP}-\tau_{START}$ between photon registration events
in start and stop detectors, tuned to indicated transitions
(start-stop order). The average excitation power 700~nW, repetition
period 6.6~ns, time bin 733~ps. Acquisition time 1~h (a, d, e) or
2.5~h (b, c, f). Single count rates 9300/s, 5800/s, 4600/s (first
detector) and 7200/s, 4500/s, 3900/s (second detector) for X, CX and
XX respectively. Solid lines represent fits of rate equation model
(see Sec.~\ref{modeldes.2}), calculated for each peak and joined
with line for better visibility.}
\end{figure*}

The results of correlation measurements involving X, CX, and XX
states from the chosen QD are presented in
Fig.~\ref{fig:histograms}.
The each of six histograms consists of distinct peaks separated by a
6.6~ns excitation repetition period. No background subtraction was
made. The signal between the peaks is negligible, indicating that
the emission from the QD is truly locked to the pulsed excitation.
The strong suppression of the central peak at $\tau=0$
(antibunching) visible in X-X, CX-CX, and XX-XX autocorrelation
histograms (Fig.~\ref{fig:histograms}(a-c)) constitutes evidence of
single photon emission. An enhancement of the central peak in XX-X
crosscorrelation (Fig.~\ref{fig:histograms}(e)) confirms the
presence of the biexciton-exciton cacscade, while its suppression in
crosscorrelations between different QD charge state transitions
(Figs.~\ref{fig:histograms}(d) and~\ref{fig:histograms}(f)) confirms
our identification, as no X-CX or XX-CX cascades are expected. These
results confirm the potential of CdTe/ZnTe QDs as sources of single
photons or photons pairs "on demand", in agreement with previous
reports.\cite{Couteau05}

Apart from the effects on the zero delay peak, all the histograms
presented show that more than one excitation period is necessary to
reach a steady state. Similar long time scale memory effects
(blinking), observed previously in autocorrelation measurements on
InAs/GaAs QDs, were interpreted recently\cite{SantoriSubm,Ulrich05}
in terms of QD charge state variation. We present here a systematic
study of these effects by measuring various types of correlations.

\begin{figure}[b]
\includegraphics[width=85mm]{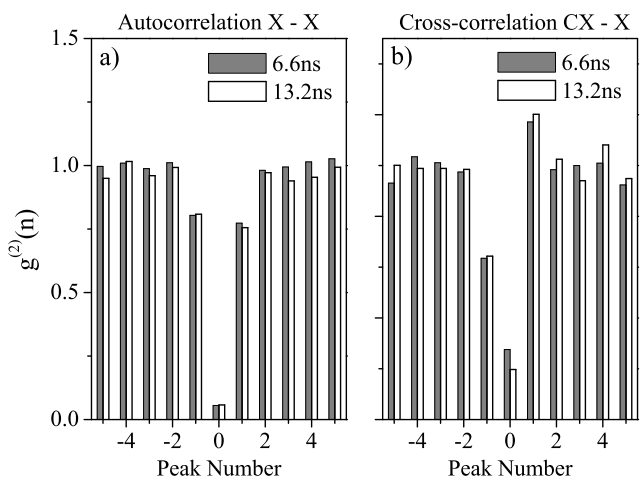}
\caption{\label{fig:comparison} X-X (a) and CX-X (b) second order
correlation functions determined with two different repetition
periods $T_{rep} = 6.6$~ns (grey bars) or 13.2~ns (white bars),
plotted as a function of peak consecutive number n~$=\tau/T_{rep}$.}
\end{figure}

To check the evolution of the QD state between the excitation pulses
we performed some of the correlation measurements at two different
repetition periods $T_{rep} = 6.6$~ns or 13~ns. We established that
the correlation functions did not depend on the time interval
between excitation pulses, as seen in Fig.~\ref{fig:comparison}. In
order to evaluate function $g^{(2)}$(n) integrated counts C(n) of
the peak number n were normalized according to the formula:
$g^{(2)}\rm{(n)}=C\rm{(n)}/(N_1N_2T_aT_{rep})$, where $N_{1,2}$ are
single counters rates, $T_a$ is total acquisition time, and
$T_{rep}$ is repetition period.\cite{Beveratos02} The negligible
contribution of background counts was not taken into account in the
calculation.

In other words the peak consecutive number n~$=~\tau/T_{rep}$ is a
good parameter to present the correlation results, rather than the
commonly used time coordinate.\cite{Ulrich05}
This indicates that the QD state remains frozen between the
excitation pulses within the accuracy of our correlation experiment,
no matter whether the correlated transitions occur in the same
(Fig.~\ref{fig:comparison}(a)) or in different
(Fig.~\ref{fig:comparison}(b)) QD charge states. Therefore in
further discussion we shall ignore in particular the long time-scale
component observed in the PL decay measurements.

\subsection{\label{sqds.Capture}Capture of single carriers by the QD}

The results of CX-X correlation measurements provide direct evidence
for the important role of single carrier capture in QD excitation.
Indeed, CX-X coincidences are registered only if the charge state of
the dot changes between the correlated emission events. If the
carriers were injected predominantly in pairs possessing no
effective charge, changes in the charge state of the dot would be
unlikely, resulting in CX-X coincidence rates much smaller than X-X
or CX-CX autocorrelation count rates, in contrast with our
experimental results. Another argument for single carrier excitation
comes from the strong asymmetry of the CX-X and XX-CX correlation
histograms (a similar result has been obtained by Kiraz et
al.\cite{Kiraz02} on InAs/GaAs QDs and Chang et al.\cite{Chang05} on
InGaAs/GaAs QDs with cw excitation). Peaks at positive (negative)
time delay in a CX-X histogram (Fig.~\ref{fig:histograms}(d))
represent detection of a neutral exciton photon after (before)
charged exciton photon. In particular reexcitation of the quantum
dot directly after the charged exciton recombination (n = 1 peak)
requires a single carrier to be trapped, while three carriers are
necessary for the opposite emission order (n = $-1$). The much
smaller probability of three trapping events following a single
pulse results in smaller intensity of the n = $-1$ peak than that of
the n = 1 one. After several pulses the QD approaches a steady state
with non-zero probabilities of finding the dot in a neutral or
charged state, resulting in an intermediate asymptotic peak
intensity. A similar explanation holds for the XX-CX
crosscorrelation (Fig.~\ref{fig:histograms}(f)).

As reported previously in respect of X emission from nonresonantly
excited III-V QDs,\cite{SantoriSubm} we also observed the
antibunching of photons emitted in n $\neq 0$ pulses: lower than
average intensities of peaks at both sides of the central one
(Fig.~\ref{fig:histograms}(a)). This indicates that excitation to X
state is less probable when the dot is empty (following an X photon
emission) than when it is in steady state condition. This difference
between the steady state and empty ground state of the dot can be
naturally explained in terms of the QD charge. In the case of XX
autocorrelation (Fig.~\ref{fig:histograms}(b)) the decrease of n~$=
\pm1$ peaks is even more pronounced. This is because as many as four
carriers are needed to repopulate the XX state. In the steady state
the finite probability of the presence of a carrier in the quantum
dot decreases the average number of carriers necessary to repopulate
the radiative state in both cases (X or XX). An explanation of the
suppression of n $=\pm1$ peaks in XX-X crosscorrelation histogram
(Fig.~\ref{fig:histograms}(e)) may be obtained in the same spirit.

The opposite conduct is observed in the charged exciton
autocorrelation: n $= \pm1$ peaks are higher than n = 0 one
(Fig.~\ref{fig:histograms}(c)). This result is explained by the fact
that after CX recombination the dot contains a single carrier, and
two more carriers (of opposite sign) must be captured to repopulate
the CX state. Peaks far from n = 0 are less intense because the
steady state involves non-zero probability of a neutral empty QD
state, requiring three carriers to enable the QD to emit another CX
photon.

In summary, we explain the observed memory effects in terms of the
QD charge state variation caused by capture of single carriers by
the QD. Within our interpretation we would not expect any memory
effects without single carrier capture: all n $\pm$ 0 peaks would
have the same intensity (since $\tau_X \ll T_{rep}$). The above
analysis allows us to make qualitative predictions concerning
correlation experiments performed under continuous wave excitation.
The suppression of X-X and XX-XX autocorrelation in the neighborhood
of zero delay should produce a broadening of the antibunching dip in
the case of cw excitation. The same suppression observed in XX-X
crosscorrelation should lead to a narrowing of the bunching peak in
cw experiments. Therefore the contribution of single carrier capture
can help to explain the unexpectedly wide antibunching dip in X-X
autocorrelation under cw excitation reported for CdTe/ZnTe
QDs,\cite{Couteau05} while no broadening was observed for the XX-X
cross-correlation peak. Furthermore, our results indicate that
lifetime determination of QD excitons based on an analysis of
correlation measurements\cite{Becher01,Persson04} may lead to
significant errors if single carrier capture is neglected.
\section{Model description of the experimental results}
\subsection{\label{modeldes.1}Rate equation model}
\begin{figure}[h]
\includegraphics[scale=0.6]{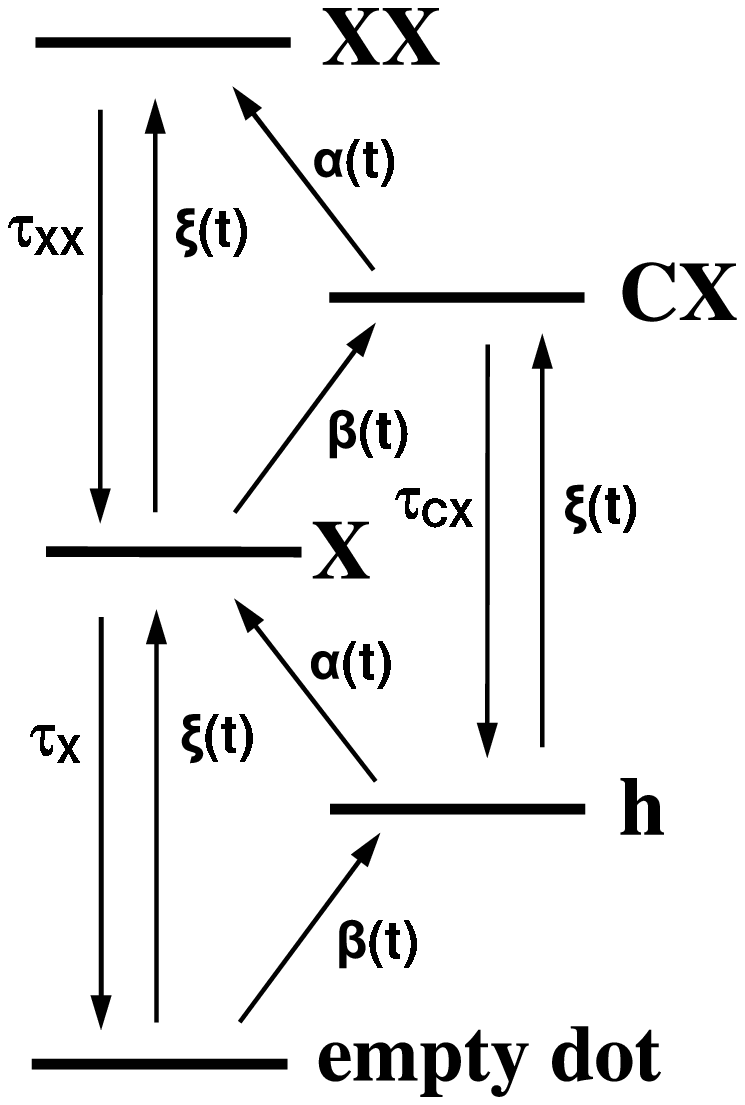}
\caption{\label{fig:ladder} Scheme of energy levels and optical
transitions considered in the model. $\alpha$(t), $\beta$(t) and
$\xi$(t) are the time dependent capture rates of the electron, hole,
and exciton respectively. Constants $\tau_X$, $\tau_{CX}$, and
$\tau_{XX}$ are decay times of respective states.}
\end{figure}

A simple model was introduced to describe our results
quantitatively. It includes occupation probabilities of five QD
states, presented as a vector ${{\bf n}({\it t}) = [p_0, p_h, p_X,
p_{CX}, p_{XX}]}$ where $p_{i}$ is the probability of finding {\it
i}-th QD state occupied at time {\it t}. Since the QD confining
potential is shallow,\cite{Piechal06} we do not include neutral or
charged excitonic levels above the biexcitonic one. Due to the low
intensity of the CX' line in the PL spectrum we also neglected
negative QD charge states. Then, possible transitions related to the
capture of a carrier or of a neutral exciton as well as to radiative
recombination are included as presented in Fig.~\ref{fig:ladder}.

Since the model deals only with level occupations, it cannot
describe any coherent effects such as, e.g., polarization
conversion.\cite{Astakhov06} It also neglects the fine structure of
the excitonic states (e.g., dark excitons). Within the assumed
approximations, the five occupation probabilities sum to unity. The
dynamics of the occupation vector can be expressed by the equation:
d{\bf n}/d{\it t} = {\it A(t)}~{\bf n}({\it t}) where {\it A(t)} is
a transition rate matrix describing radiative decays and excitation
processes:
\begin{widetext}
\begin{equation}
A(t)=\left[
\begin{array}{ccccc}
-\alpha(t) -\xi(t) & 0 & 1/\tau_X & 0 & 0 \\
\alpha(t) & -\beta(t) - \xi(t) & 0 & 1/\tau_{CX} & 0 \\
\xi(t) & \beta(t) & -\alpha(t)-\xi(t)-1/\tau_X & 0 & 1/\tau_{XX} \\
0 & \xi(t) & \alpha(t) & -\beta(t)-1/\tau_{CX} & 0 \\
0 & 0 & \xi(t) & \beta(t) & -1/\tau_{XX}
\end{array}
\right]
\end{equation}
\end{widetext}
The important role of trapping of single carriers has already been
discussed above. We also include the possibility of excitation by
trapping free excitons which (as follows from our simulations) is
needed to achieve the observed quadratic excitation-power dependence
of XX line intensity. The characteristic timescales of the electron,
hole and exciton trapping (which may be different from one another)
are larger than the laser pulse width because of relaxation
processes within the barrier material. With the small outer barrier
thickness of 50~nm we expect the trapping rates to rise almost
instantaneously and to decay with characteristic times of the order
of the barrier exciton lifetime (Sec.~\ref{sec:Sample}). Therefore
we assumed as a starting point a common exponential time dependence
of the three excitation rates, with a decay time of $\tau_{exc} =
20$~ps. As we know the three radiative lifetimes $\tau_X$,
$\tau_{CX}$ and $\tau_{XX}$ from independent experiments, the only
free parameters of the model remain constants $\alpha$, $\beta$ and
$\xi$ understood as time-integrated respective capture rates, e.g.,
$\alpha(t) = (\alpha/\tau_{exc})exp(-t/\tau_{exc})\theta(t)$, where
the Heaviside function $\theta(t)$ equals to 0 for $t < 0$ and to 1
for $t \geq 0$. We also allow for an arbitrary coefficient between
the computed and measured count rates to account for (unknown)
photon detection efficiency.

We integrated the rate equations numerically to compute {\bf n}({\it
t}). Integration over time of $I(t)=p_X/\tau_X$, $p_{CX}/\tau_{CX}$,
$p_{XX}/\tau_{XX}$ gives PL intensities (per excitation pulse) for
X, CX or XX respectively. The initial state is defined by a steady
state or by a defined transition observed in the correlation
experiment.

If we choose the initial point for integration to meet the steady
state condition {\bf n}({\it t}) = {\bf n}$(t + T_{rep})$, we get
relative line intensities in the PL spectrum. As far as the
correlation experiment is concerned, the first photon unequivocally
defines {\bf n}(0) (e.g., if we detect a photon from the
recombination of CX, we know the dot is occupied by a hole) and
therefore we can calculate the expected PL intensity for the following
$T_{rep}$ period. The correlation function is proportional to
this expected PL multiplied by the average PL intensity of the line
of the first photon. The calculation of the correlation function
$g^{(2)}(\tau)$ for the central peak is slightly more complicated
and can be written as:
\begin{widetext}
\begin{subequations}
\begin{equation}
g^{(2)}(\tau = 0) = \frac{1}{I_{AB}}\int_{0}^{T_{rep}}\!\!\!\int_{\tau'}^{T_{rep}}\left[I_{A,s}(\tau')I_{B,a}(\tau'') + I_{B,s}(\tau')I_{A,b}(\tau'')\right]\ {\rm d}\tau''\ {\rm d}\tau',
\end{equation}
\begin{equation}
I_{AB} = \int_{0}^{T_{rep}}I_{A,s}(\tau')\ {\rm d}\tau' \int_{0}^{T_{rep}}I_{B,s}(\tau')\ {\rm d}\tau',
\end{equation}
\end{subequations}
\end{widetext}
where in the example case of X-CX crosscorrelation "A" denotes the X
state, "a" denotes the QD state after X photon emission (empty dot),
B denotes the CX state, "b" denotes the QD state after CX photon
emission (a single hole present), and "s" denotes the steady state.
$I_{K,m}(t)$ is time dependence for a rate of emission from QD state
denoted as "K" when starting from the state denoted as "m".

\subsection{\label{modeldes.2}Description of the experimental results}
The results of fitting the rate equation model to experimental data
are presented in Figs.~\ref{fig:PowerDep} and \ref{fig:histograms}.
The fitting procedure takes into account the results of PL intensity
power dependence and the complete set of auto- and crosscorrelation
measurements simultaneously. As seen in Fig.~\ref{fig:histograms},
the model describes the results of correlation measurements with
excellent accuracy. The shapes of the correlation histograms, i.e.
long timescale bunching and antibunching features are repeated by
the model with high fidelity. The calculated number of counts in the
zero delay peak on each histogram in Fig.~\ref{fig:histograms}
agrees with values obtained in experiment. The same set of fitting
parameters allows the slopes and magnitude of X, CX and XX emission
intensities to be described, plotted as a function of excitation
power in Fig.~\ref{fig:PowerDep}. However, the region of strong,
saturating excitation powers is not described adequately by the
model (not shown). This aspect requires further study.

As a result of the fitting procedure, integrated capture rates per
pulse for electron, hole and exciton were obtained: $\alpha = 0.80$,
$\beta= 0.86$, and $\xi = 0.26$, respectively. We see that the
capture of a single carrier is more than three times more probable
than capture of an electron-hole pair. This supports our
interpretation (Sec.~\ref{sqds.Capture}) of the results of the auto-
and crosscorrelation experiment indicating the role of single
carrier capture processes in QD excitation.

The systematic simulations have convinced us that the integrated
capture rates are the most important parameters influencing the PL
intensity and correlation results. The detailed temporal
distribution of the capture processes is less important, provided
that their characteristic times are significantly lower than the QD
PL decay times or the excitation repetition period. To study these
processes in detail, different tools should be applied, such as
pump-probe absorption measurements or Correlation Excitation
Spectroscopy.

\section{Conclusions}
We demonstrate the utility of photon correlation measurements to
study non-resonant excitation mechanisms of semiconductor quantum
dots. A qualitative analysis indicates the important role of single
carrier capture processes and leads to an explanation of the
unusually wide antibunching dip observed in previously reported
autocorrelation measurements under cw excitation. The rate equation
model introduced allowed us to describe a complete set of
correlation and PL intensity results and to obtain quantitative
information on the trapping rates of electrons, holes, and excitons
by the quantum dot.
\begin{acknowledgments}
This work was supported by the Polish Committee for Scientific
Research (grants 2P03B 002 25, 2P03B 015 25, and
PBZ-KBN-044/P03/2001), and the Polonium program.
\end{acknowledgments}


\end{document}